\documentclass[twocolumn,epsfig,superscriptaddress]{revtex4}

\usepackage{graphicx}
\usepackage{amssymb}
\usepackage{amsmath}
\usepackage{color}
\usepackage{bm}
\usepackage{physics}

\usepackage{changes}

\makeatletter
\let\Changes@Markup@Deleted\@gobble
\makeatother

\DeclareMathOperator\arccosh{arccosh}

\newcommand{\figsizeone}{0.4}
\newcommand{\figsizetwo}{0.45}

\begin{document}

\title{Topological edge states in bowtie ladders with different cutting edges}

\author{Jung-Wan Ryu}
\affiliation{Center for Theoretical Physics of Complex Systems, Institute for Basic Science, Daejeon 34051, Republic of Korea}
\author{Sungjong Woo}
\affiliation{Center for Theoretical Physics of Complex Systems, Institute for Basic Science, Daejeon 34051, Republic of Korea}
\author{Nojoon Myoung}
\affiliation{Department of Physics Education, Chosun University, Gwangju 61452, Republic of Korea}
\author{Hee Chul Park}
\affiliation{Center for Theoretical Physics of Complex Systems, Institute for Basic Science, Daejeon 34051, Republic of Korea}

\begin{abstract}
We have studied topological edge states in bowtie ladders with various edge truncations. The symmetric bowtie ladder, which comprises two trivial Su--Schrieffer--Heeger (SSH) lattices, exhibits an insulator-metal transition with trivial insulating states. On the other hand, the lattice can be transformed into an extended SSH lattice depending on the edge shapes with non-trivial insulating states in that the winding number is non-zero. The winding numbers are permutationally designated in the phase diagram depending on the choice of unit cell. The topological edge states are affected by the shape of the edge and the corresponding winding number. We also studied general bowtie ladder models with richer phase diagrams using the characteristics of the localization length of the edge states showing state bifurcation.
\end{abstract}

\maketitle

\section{Introduction}

The last decade has witnessed the discovery of topological materials with which unconventional models and materials have been theoretically proposed and experimentally revealed \cite{Hasan, Xiao}. Research on topological and artificial models has become increasingly pervasive due to the possibility of non-triviality. 
In condensed matter physics, topological materials are defined by the winding number/topological charge in bulk, the so-called invariant of the topological band, and the corresponding edge states imposed by boundary conditions \cite{Fradkin}.
The relationship between the winding number of bulk lattices and the edge states of finite lattices is fundamental to describe the characteristics of topological insulators \cite{Del11, Asb16}. The bulk-boundary correspondence represents the fulcrum of the strong evidence behind topological materials. Given a unit cell, the winding number is introduced with a single period of Bloch wave vector $\bold{k}$ by using the Bloch Hamiltonian in the Brillouin zone. Its wave function provides a Berry connection that is defined by the inner product of the wave function and its partial derivative. The integral of the Berry connection with respect to the wave vector over the whole Brillouin zone is a Zak phase or Chern number depending on dimensionality \cite{Schnyder, Xiao08}. The Zak phase, quantized under chiral symmetry in one dimension, is related to the winding number corresponding to modern polarization \cite{Ryu02, Mer76, Zak89, King93}. The non-triviality of those quantities is complementary to the existence of edge states at the edges of a proper shape \cite{Fradkin}. 

The prototypical model to exhibit bulk-boundary correspondence is Su--Schriefer--Heeger (SSH) model \cite{Su79}. Although this model provides us with a basic intuition of topological phases, the choice of unit cell is allowed to contain ambiguity for the topological invariants in one-dimensional systems \cite{Ryu02, Stone84, Ata13, Jun17}. 
Thus, the relative differences in winding numbers are important to designate topological non-triviality. In addition, different shapes of the edges have been shown to lead to different topological phases according to the unit cell in the SSH model, carbon nanotubes, and graphene \cite{Ryu02, Del11, Jun17, Jun18}. Topological defects, moreover, introduce topological edge states in trivial insulating lattices, even breaking the symmetry \cite{Lee18, Han19}.

In this work, we study the topological properties of a bowtie ladder lattice with direct and cross inter-chain hoppings and without intra-chain hoppings, depending on the shape of the cutting edge. The bowtie ladder, which can be considered as coupled SSH chains \cite{Li17}, has attracted interest as a canonical ladder geometry from which any other topological ladder model can be obtained by unitary transformation \cite{Car19}. It is known that a bowtie ladder with symmetric cross inter-chain hopping exhibits a transition between gapped and gapless states with a trivial topological invariant, while a bowtie ladder with asymmetric cross inter-chain hopping shows a greater variety of topological phases \cite{Li17}. 
We demonstrate non-trivial bulk topology and corresponding edge states due to ambiguity in the selection of unit cells in a bulk bowtie ladder with symmetric cross inter-chain hoppings, which until now have been considered topologically trivial. 
In addition, we explore the relation between the topological invariants of bowtie ladders, i.e., bulk winding numbers, as determined by different unit cells in cases with symmetric and asymmetric cross inter-chain hoppings.

\section{Hamiltonians of Bowtie ladders}

\begin{figure}
\begin{center}
\includegraphics[width=\figsizeone\textwidth]{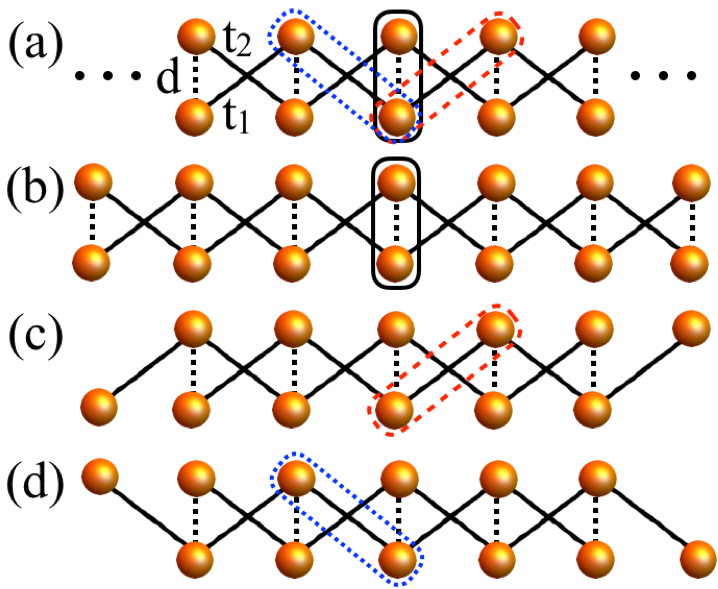}
\caption{(color online). (a) Bowtie ladder with direct inter-chain hopping amplitude $d$ (black dotted lines) and cross inter-chain hopping amplitudes $t_{1}$ and $t_{2}$ (solid lines). The system consists of two sublattices $A$ and $B$, indicated by upper and lower spheres, respectively. The black ($V$), red dashed ($O^{+}$), and blue dashed ($O^{-}$) boxes represent unit cells with vertically, positively, and negatively rotated obliquely arranged sublattices, respectively. (b--d) Finite-sized bowtie ladders with a vertical edge shape (b), and positively (c) and negatively (d) rotated oblique edge shapes, corresponding to the unit cells with vertically and obliquely arranged sublattices.
}
\label{fig1}
\end{center}
\end{figure}

The bulk states of a periodic system are determined by crystal momentum $k$. The quantum number $k$ is replaced with $k=-i\frac{\partial}{\partial x}$ via Peierls substitution giving trial wave function $\psi=e^{ik}$. Counting the winding number is equivalent to the Zak phase for the Hamiltonian $H(k)$, and quantized winding numbers give information on the edge states \cite{Zak89}. On the other hand, secular equations $\det[H(k)-\varepsilon_k]=0$ give the eigenvalues $\varepsilon_k$ with eigenfunctions for the Hamiltonian. These bulk states, however, cannot describe localized states at the interface or on the edges in a finite lattice, which are delocalized through the entire lattice. 
We try a decaying wave function with oscillation for the ansatz as a generalized Bloch factor, in order to find the edge states. The ansatz contains the complex momentum $k=q+i\kappa$, where $q$ is a wave number and $\kappa$ is the spatial decay rate \cite{Stone84, Bin08, Xiao14}. The secular equation with the complex momentum determines $q$ and the localization length $\xi=1/\kappa$ with corresponding eigenvalue \cite{Xiao14, Chao19, Lix01}.

We introduce a general bowtie ladder, as shown in Fig. \ref{fig1} (a), and clarify the topological edge states on the lattices.
The states on the sublattices $A$ and $B$ in unit cell $m$ are denoted by $|m,A\rangle$ and $|m,B\rangle$, respectively.
To represent a bipartite Hamiltonian, it practical to separate the external degrees of freedom (unit cell index $m$) from the internal degrees of freedom (sublattice indices $A$ and $B$). We can use a tensor product basis, $|m,\alpha\rangle \rightarrow |m\rangle \otimes |\alpha\rangle$ and $H=H_{external} \otimes H_{internal}$, with $\alpha \in {A,B}$.
The Hamiltonian $H$ conserves the combination of the time reversal $\mathcal{T}=\mathcal{K}$ and inversion $\mathcal{M}=\sigma_x$ symmetries $[H,\mathcal{TM}]=0$ and satisfies anti-commutation relations with charge conjugate $\mathcal{C}=\sigma_z\mathcal{K}$ and chiral symmetry $\mathcal{S}=\sigma_z$, where $\mathcal{K}$ is complex conjugation operator.
The Hamiltonians based on a vertical unit cell and rotated oblique unit cells can be written as follows,
\begin{eqnarray}
H^{\mu}=\sum^{}_{m} h^{\mu}_{m}\otimes \frac{\sigma_{x} + i \sigma_{y}}{2}+ H.C.,
\end{eqnarray}
where $\mu=(-,0,+)$ indicates the unit cell and $h^{\mu}_{m}$ is the hopping sectors of the whole Hamiltonian which are defined by 
\begin{eqnarray}
\begin{matrix}
    h^{0}_{m}=d |m\rangle \langle m|+t_1 |m+1 \rangle \langle m|+t_2 |m-1 \rangle \langle m|\\
    h^{+}_{m}=t_1 |m\rangle \langle m|+d |m+1 \rangle \langle m|+t_2 |m+2 \rangle \langle m|\\
    h^{-}_{m}=t_2 |m\rangle \langle m|+d |m-1 \rangle \langle m|+t_1 |m-2 \rangle \langle m|,
\end{matrix}
\end{eqnarray}
respectively, where $d$ is a direct inter-chain hopping parameter, $t_{1}$ and $t_{2}$ are cross inter-chain hoppings [as shown in Fig.~\ref{fig1} (a)], and $\bold{\sigma}$ is the Pauli matrix. The Hamiltonian based on a negatively rotated oblique unit cell can be obtained by changing $t_{1,2}$ into $t_{2,1}$ and using a time reversal operator ($k\rightarrow-k$).
The two-band Hamiltonian $H(k)$ in momentum space (i.e., a model with two internal states per unit cell), reads
\begin{equation}
    H(k) = d_{x}(k) \sigma_{x} + d_{y}(k) \sigma_{y} + d_{z}(k) \sigma_{z} = \bold{d} (k) \cdot \bold{\sigma}.
\end{equation}
The real numbers $d_{x,y,z} \in \mathbb{R}$, the components of the k-dependent three-dimensional vector $\bold{d}(k)$, and the internal structure of the eigenstates with momentum $k$ are all given by the direction in which vector $\bold{d}(k)$ points.

\section{Bowtie ladders with symmetric cross inter-chain hoppings}

Let us start with a Hamiltonian of a bowtie ladder with symmetric cross inter-chain hoppings, $t = t_{1} = t_{2}$, using a vertical unit cell with given $k$ in momentum space $h^{0}(k)=\bold{d^{0}}(k)\cdot\bold{\sigma}$, where the vectors $d^{0}_x(k)=1+2t\cos{k}$ and $d^{0}_y(k)=d^{0}_z(k)=0$ with a cross inter-chain hopping parameter $t$. All parameters are normalized by direct inter-chain hopping amplitude $d$ throughout this paper. This symmetric bowtie ladder lattice can be considered as a combination of two trivial SSH models that shows a insulator-metal transition.
As the wave number runs through the Brillouin zone, $k = 0 \rightarrow 2 \pi$, the path that the endpoint of the vector $\bold{d}(k)$ traces out is a line on the ($d_{x},d_{y}$) plane due to the periodicity of the bulk momentum-space Hamiltonian, and it needs to avoid the origin to describe an insulator or conversely needs to meet the origin to do so a conductor. A bowtie ladder described by a vertical unit cell can thus become a topologically trivial insulator or conductor depending on the parameters. This can be explained via the quantum phase transition associated with the spontaneous symmetry breaking between gapped and gapless states with a trivial topological invariant, i.e., a zero winding number, in coupled SSH chains \cite{Li17}.
From the view point of the winding number, it shows a metal-insulator transition at $t=1/2$ with the metal showing a gapless energy band under $t\geq 1/2$ and the insulator showing a trivial, gapped energy band under $t<1/2$.
Solutions of the secular equation, however, provide the zero energy condition of this system, $X^2+X/t+1/4t^2=0$, where $X = \cos{k}$. Considering a finite boundary condition with complex momentum $k=q+i\kappa$, the solution is $\cosh{\kappa}=1/2t$ when $q=\pi$. This means that a localized state exists in the trivial insulating regime having a finite localization length, $\xi=1/\kappa=1/\arccosh{(1/2t)}$, which seems to be discrepant with the bulk-boundary correspondence.

\begin{figure}
\begin{center}
\includegraphics[width=\figsizetwo\textwidth]{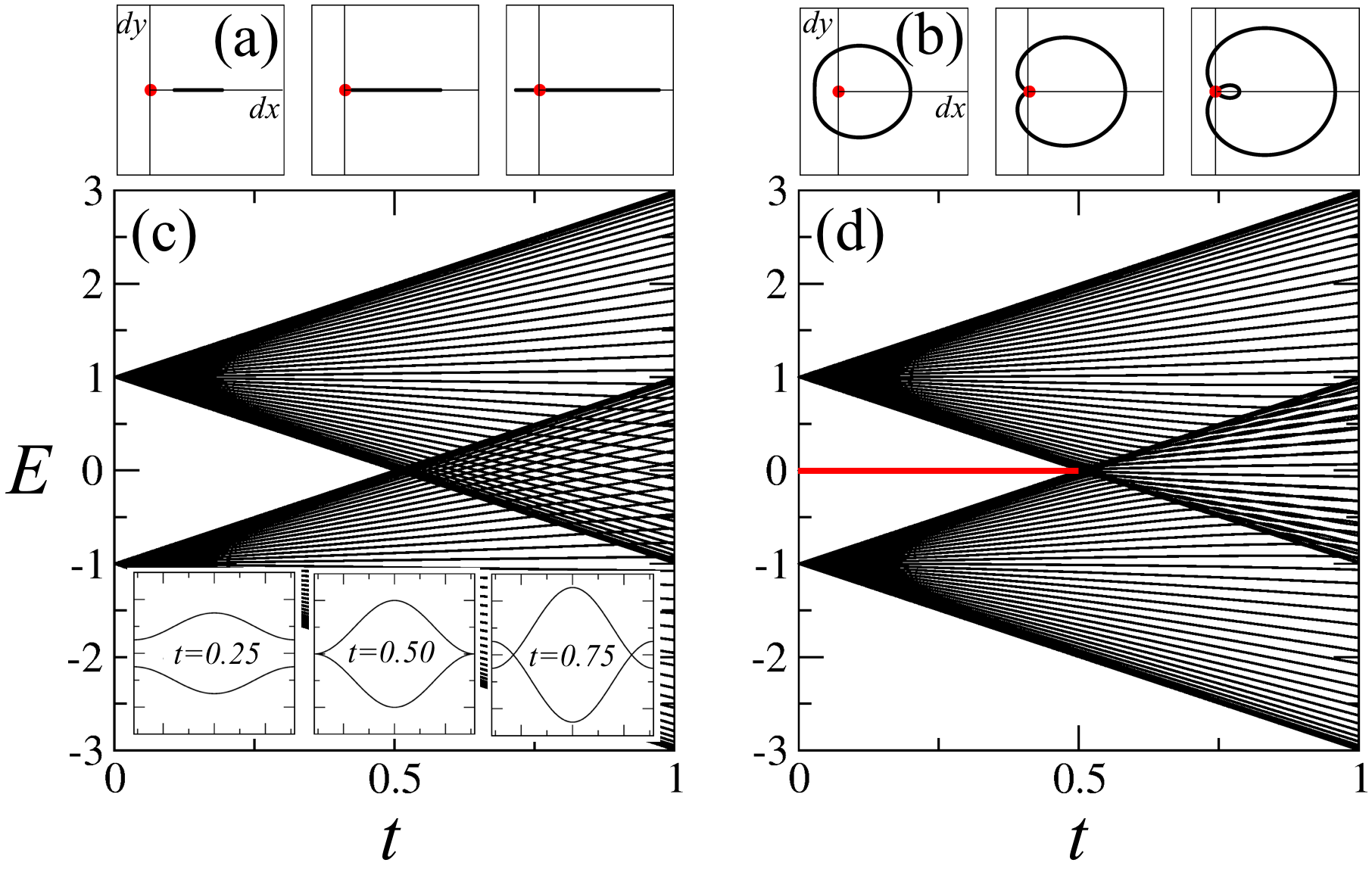}
\caption{(color online). (a,b) Paths of the vector $\bold{d}(k)$ on the $(d_x, d_y)$ plane with respect to the Brillouin zone, $k = 0 \rightarrow 2 \pi$, of bowtie ladders with vertical (a) and oblique (b) unit cells corresponding to $t = 0.25$, $0.5$, and $0.75$, respectively. The red dots represent the origin. (c,d) Eigenenergies as a function of $t$ in bowtie ladders in which cross inter-chain hoppings are symmetric with vertical (c) and oblique (d) cutting edges. The red line represents zero energy edge states.
The insets in (c) are the band structures for $t = 0.25$, $0.5$, and $0.75$.
}
\label{fig2}
\end{center}
\end{figure}

This discrepancy implies that there are various topological phases in the lattices containing multiple sites in a unit cell.
There are three ways to take unit cells in a bowtie ladder based on three different orientations: negative oblique ($O^{-}$), vertical ($V$), and positive oblique ($O^{+}$), as shown in Fig. \ref{fig1} (a).  
For the oblique orientations, the constituent vectors of the Hamiltonians having oblique unit cells are $d_{x}^{\pm}(k)=\cos{k}+t(\cos{2k}+1)$, $d_{y}^{\pm}(k)=\pm(\sin{k}+t\sin{2 k})$, and $d_{z}^{\pm}(k)=0$, as distinguished from the case of the vertical unit cell. All of the Hamiltonians can be diagonalized as
\begin{equation}
\label{ek}
    E_{k} = \pm \sqrt{1 + 2 t^2 + 4 t \cos{k} + 2 t^2 \cos{2 k}}.
\end{equation}
We show the dispersion relations for the three choices of parameters in Fig.~\ref{fig2} (c). The bowtie ladder model describes a conductor when $t>0.5$, while there is an energy band gap separating the lower filled band from the upper empty band when $t<0.5$. A marginal case occurs at $t = 0.5$.

Although dispersion relations are useful to check off a number of the physical properties of the bulk of the system, there is also important system information that they do not reveal. Here, the corresponding winding numbers are $\bar{\nu}=\{-1,~0,~1\}$ for the unit cells of the negative oblique, vertical, and positive oblique orientations, respectively, in the insulating phase $t<1/2$. The bowtie ladders described by an oblique unit cell reach an insulating phase with a non-trivial topological invariant, or corresponding winding number $\nu^{\pm} = \pm 1$ as shown in Fig.~\ref{fig2} (b). In bowtie ladders with oblique unit cells, there is a quantum phase transition associated with non-zero winding numbers, unlike the quantum phase transition associated with the zero winding number in bowtie ladders with a vertical unit cell.

For finite systems, we can truncate both edges according to the shapes of the unit cells, as shown in Fig. \ref{fig1} (b--d). The spectrum of an open bowtie ladder of $N = 40$ unit cells evolves as we continuously turn on the hopping amplitude $t$ with the prepared lattices. Independent of the choice of unit cell, the eigenenergies always appear as a pair, i.e., $E$ and $-E$, because of particle-hole symmetry.
According to the bulk-edge correspondence \cite{Hasan, Xiao}, there should be gapless edge states when the bulk states are topologically non-trivial for the oblique shapes, while there is no edge state for the vertical shape. The spectra in Fig.~\ref{fig2} (d) reveal that the energies of the edge states in the bowtie ladders with oblique cutting edges remain very close to zero energy, while there are no edge states in the case with a vertical cutting edge in Fig.~\ref{fig2} (c). The wavefunctions of a pair of almost-zero-energy edge states have to be exponentially localized at both the left and right edges, because the zero energy is in the bulk band gap. The zero energy edge states in the case of insulators show a link between the bulk winding number and the absence or presence of edge states, known as bulk-boundary correspondence. 
Solutions of the secular equations $H(k)-\varepsilon_k|_{\varepsilon_k=0}$ with complex momentum $k=q+i\kappa$ provide the inverse localization length at zero energy,
\begin{equation}
\label{ll_sym}
    \kappa = \arccosh{\left(\frac{1}{2 t}\right)}.
\end{equation}
The localization length of the zero energy edge states is $\xi = 1 / \kappa$ \cite{Del11}, which corresponds to the localization length of the zero energy states when $0 < t < 0.5$ in Fig.~\ref{fig2} (d).

\section{Generalized bowtie ladders with asymmetric cross inter-chain hoppings}

\begin{figure}
\begin{center}
\includegraphics[width=\figsizeone\textwidth]{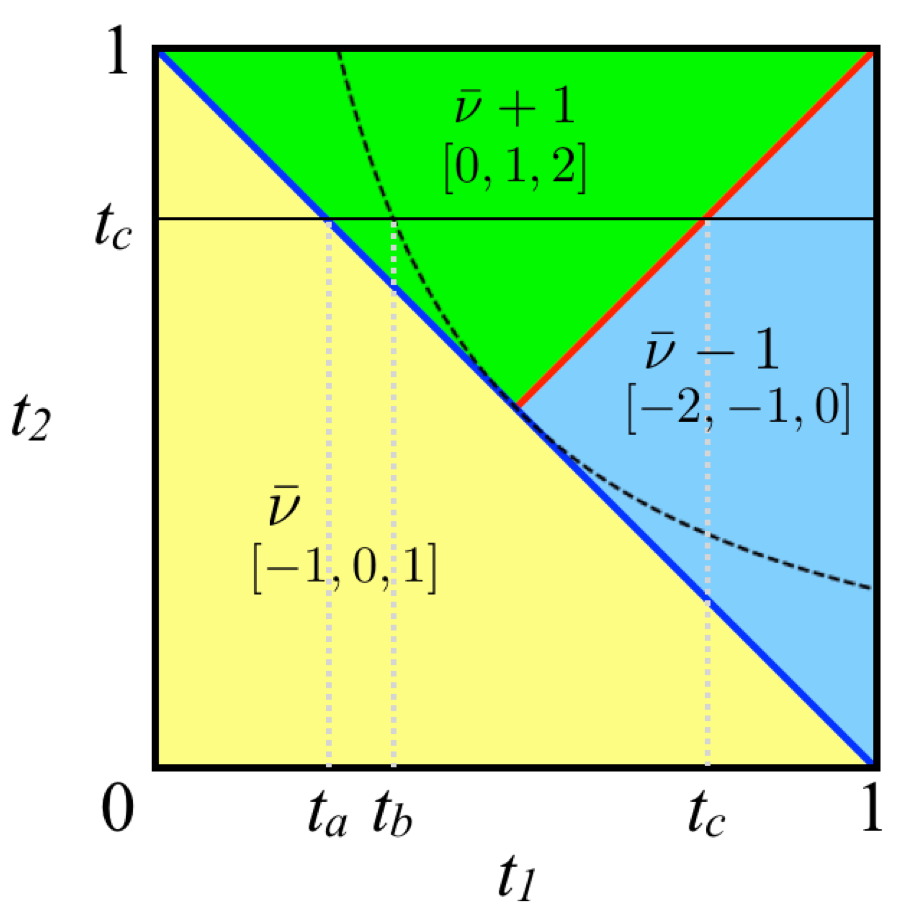}
\caption{(color online). Phase diagram of a bowtie ladder on the ($t_1$, $t_2$) plane. The winding numbers [$\nu^{-}$, $\nu^{0}$, $\nu^{+}$] as a function of $(t_{1}, t_{2})$ correspond to $2 |\nu|$ edge states. The numbers [$\nu^{-}$, $\nu^{0}$, $\nu^{+}$] represent the winding number in bowtie ladders with negatively rotated oblique, vertical, and positively rotated oblique cutting edges, respectively. The red and blue lines represent conductors, and the black dashed line denotes $4 t_{1} t_{2} = 1$ where the localization lengths bifurcate. The black solid line indicates $t_{2} = 0.75$.
}
\label{fig3}
\end{center}
\end{figure}

\begin{figure}
\begin{center}
\includegraphics[width=\figsizetwo\textwidth]{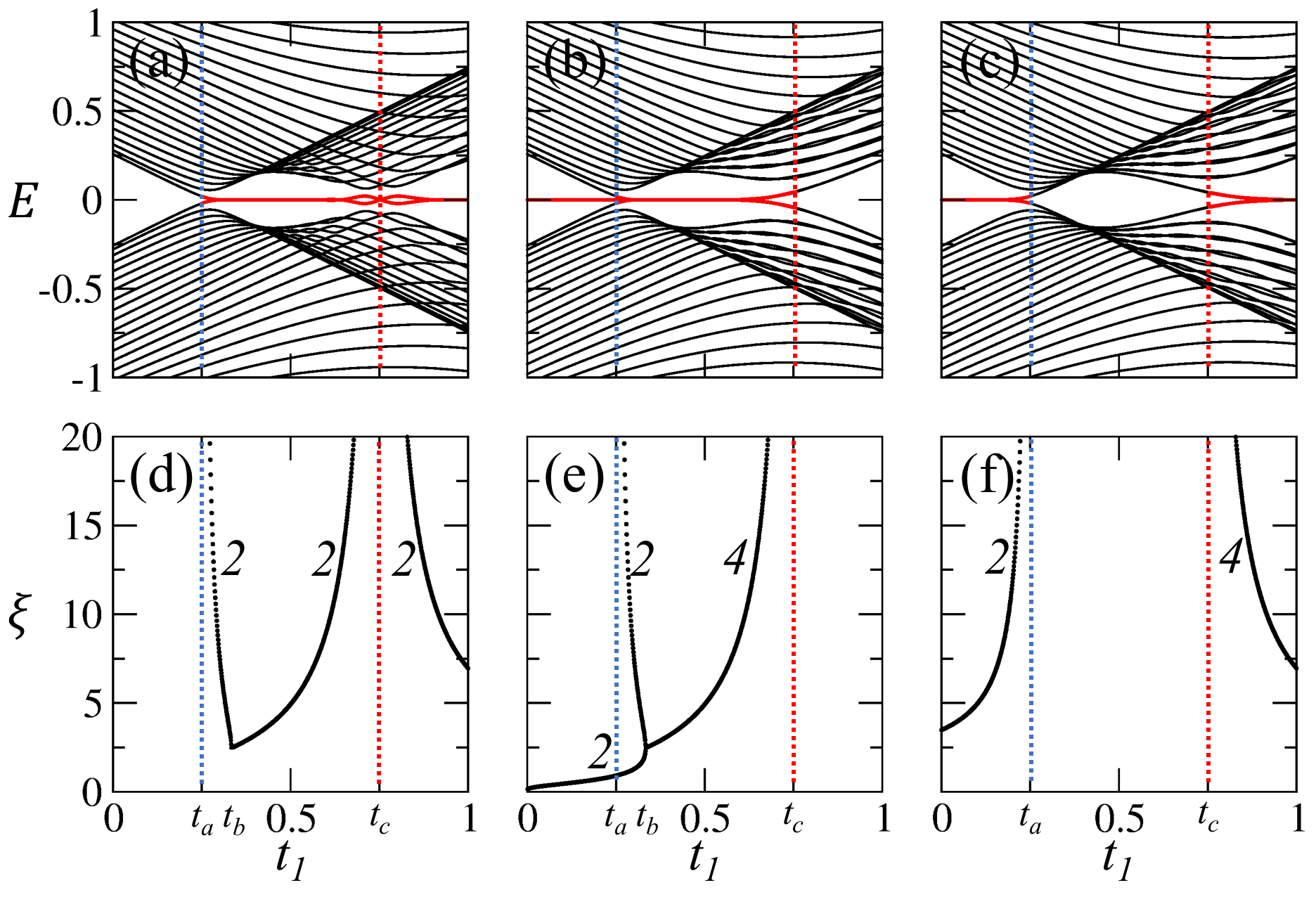}
\caption{(color online). Eigenenergies as a function of $t_{1}$ in bowtie ladders with (a) vertical, (b) positively rotated oblique, and (c) negatively rotated oblique cutting edges when $t_{2} = 0.75$. The red lines represent localized edge states.
The black dots represent the localization lengths as a function of $t_{1}$ in bowtie ladders with (d) vertical, (e) positively rotated oblique, and (f) negatively rotated oblique cutting edges when $t_{2} = 0.75$. The blue and red dotted lines represent $t_{1} = 0.25$ and $t_{1} = 0.75$, respectively, where the systems are conductors. The numbers represent the numbers of corresponding edge states.
}
\label{fig4}
\end{center}
\end{figure}

Next, we consider the Hamiltonian of a bowtie ladder with asymmetric cross inter-chain hoppings. The constituent vectors of the Hamiltonians in the cases of vertical, positive oblique, and negative oblique orientations are
\begin{eqnarray}
\label{dx_v}
    d_{x}^{0}(k) &=& (t_{1} + t_{2}) \cos{k} + 1, \\\nonumber
    d_{y}^{0}(k) &=& -(t_{1} - t_{2}) \sin{k},\\
\label{dx_p}
    d_{x}^{+}(k) &=& \cos{k} + t_{2} \cos{2 k} + t_{1}, \\\nonumber
    d_{y}^{+}(k) &=& \sin{k} + t_{2} \sin{2 k},\\
\label{dx_m}
    d_{x}^{-}(k) &=& \cos{k} + t_{1} \cos{2 k} + t_{2}, \\\nonumber
    d_{y}^{-}(k) &=& -\sin{k} - t_{1} \sin{2 k},\nonumber
\end{eqnarray}
respectively.
The phase diagram of a bowtie ladder with the vertical unit cell corresponding to a vertical cutting edge is well known \cite{Li17}, as shown in Fig.~\ref{fig3}. The topological properties in each phases can be characterized by the loop of the curve obtained from Eq.~(\ref{dx_v}) on the $(d_{x}, d_{y})$ plane \cite{Zha15}. When $(t_{1}+t_{2}) < 1$, the system describes inter-chain dimerization resulting in an insulator with trivial topological invariants, corresponding to winding number $\nu^{0} = 0$. Therefore, there are no edge states. When $(t_{1}+t_{2}) > 1$, the system describes a conductor if $t_{1} = t_{2}$ and thus the winding number $\nu^{0}$ is undefined, while the system describes an insulator with non-trivial topological invariants corresponding to winding numbers $\nu^{0} = \pm 1$ for $t_{1} < t_{2}$ and $t_{1} > t_{2}$, respectively. In this case, two zero energy edge states emerge. Figure~\ref{fig4} (a) shows the eigenenergy spectra as $t_{1}$ increases with fixed $t_{2}$. At $t_{2} = t_{c}$ where $0.5 < t_{c} < 1$, there are no zero energy edge states when $t_{1} < t_{a} = 1 - t_{c}$, while a pair of zero energy edge states with positive and negative winding numbers appear when $t_{a} < t_{1} < t_{c}$ and $t_{1} > t_{c}$, respectively, in Fig.~\ref{fig4} (a). 

In the case of a bowtie ladder with an oblique cutting edge corresponding to an oblique unit cell, the critical lines between different physical phases are the same as the case with the vertical cutting edge. However, the physical phases characterized by the winding numbers are different. The winding numbers of a bowtie ladder with positively and negatively rotated unit cells is 1 greater or less than those of a bowtie ladder with a vertical unit cell, respectively. Considering the positively rotated cutting edge, when $(t_{1}+t_{2}) < 1$, the system describes an insulator with non-trivial topological invariants, corresponding to winding number $\nu^{+} = 1$. Thus, there is a pair of topologically protected edge states. When $(t_{1}+t_{2}) > 1$, the system describes a conductor if $t_{1} = t_{2}$, and thus the winding number $\nu^{+}$ is undefined. This system describes an insulator with trivial topological invariants, corresponding to winding numbers $\nu^{+} = 0$ if $t_{1} > t_{2}$. However, the system describes an insulator with non-trivial topological invariants, corresponding to winding number $\nu^{+} = 2$ if $t_{1} < t_{2}$, and accordingly, there are two pairs of edge states. At $t_{2} = t_{c}$ where $0.5 < t_{c} < 1$, there is a pair of zero energy edge states when $t_{1} < t_{a}$, two pairs of zero energy edge states when $t_{a} < t_{1} < t_{c}$, and no edge states when $t_{1} > t_{c}$, as in Fig.~\ref{fig4} (b).

The topology of the energy bands in each area can be characterized by a loop in the auxiliary space $(d_{x}(k),~d_{y}(k))$, $k\in [-\pi,~\pi]$. The feature of the quantum phase is characterized by the topology of the loop. The winding number of the loop around the origin of the ($d_{x}$, $d_{y}$) plane is defined as
\begin{equation}
    \nu = \frac{1}{2 \pi} \int_c{\left(d_{x} \text{d}d_{y} - d_{y} \text{d}d_{x}\right)/r^2},
\end{equation}
where $r^2 = d_{x}^2 + d_{y}^2$. A straightforward derivation from the above definition yields $\nu^{0} = 0$ when $t_{1} + t_{2} < 1$ for the bowtie ladder with a vertical cutting edge. Otherwise, $\nu^{0} = \mathrm{sgn}(t_{2}-t_{1})$. As a result, the winding numbers between bowtie ladders with vertical and oblique cutting edges show the relation
\begin{equation}
\label{rwn}
    \nu^{\pm} = \nu^{0} \pm 1.
\end{equation}
This means that the topological properties of bowtie ladders depend on the cutting edges. It is noted that the winding number of each configuration is a gauge-dependent quantity, i.e., it depends on the choice of unit cell associated with the cutting edge shape. However, the difference between the winding numbers of the two configurations is uniquely defined. 

We consider the properties of the zero energy localized states characterized by localization length.
The localization lengths of the edge states can be also obtained from the secular equations with the complex momentum $k=q+i\kappa$ used in the previous section. In general, two kinds of solutions are possible: one in which the radicand is positive and one in which it is negative, according to the competition between inter- and intra-cell hopping. Under the condition $t_{2}\leq 1/4 t_{1}$ as shown by the black dashed line in Fig.~\ref{fig3}, the localization length is written as follows,
\begin{eqnarray}
\xi_{\pm}=1/\arccosh{\left(\frac{t_+ \pm t_-\sqrt{1-4t_1t_2}}{4t_1t_2}\right)},
\end{eqnarray}
where $t_{\pm}=t_1\pm t_2$ and the wave vector is constant $q=\pi$ which means that the wave function is monotonically decaying.
Under the other condition, $t_{2} > 1/4 t_{1}$, the localization length is found as follows,
\begin{eqnarray}
\xi= 2/\abs{\log{(t_1/t_2)}},
\label{eq:2log}
\end{eqnarray}
where the localized edge states are spatially oscillating with the wave vector $q=\arctan{\sqrt{4t_1t_2-1}}$. We can see a transition through the bifurcation of the localization length at $t_{2} = t_{b}$ in Fig. \ref{fig4} (e).

Figure \ref{fig4} (d) shows the localization length of the localized states for the vertical cutting edge. The localization length of a pair of edge states with $\nu^{0}=1$ decays from infinity to a finite value in $t_a<t_{1}<t_{b}$ and then increases to infinity in $t_b<t_{1}<t_{c}$ as $t_{1}$ increases. The localization length of a pair of edge states with $\nu^{0}=-1$ is diminished as a function of $t_{1}$ in $t_{1}>t_{c}$.

In the case of the positive oblique cutting edge, as shown in Fig. \ref{fig4} (e), the localization length of a pair of edge states increases with $\nu^{+}=1$ in $0<t_{1}<t_{a}$. The condition provides that the localization length is $\xi_-$. The localization length approaches $\xi_-\rightarrow 0$ as $t_{1}\rightarrow 0$. It is noted that there exist localized states with a finite localization length, $\xi_+=(1+(2t_1-1)^2)/4t_1(1-t_1)$, although the bulk gap is closed at the blue transition line $t_1 = t_a$.
There are two pairs localized edge states with $\nu^{+}=2$ in $t_{a} < t_{1} < t_{c}$, showing different localization lengths in $t_{1} < t_{b}$ and the same localization lengths in $t_{b} < t_{1} < t_{c}$. For the former case, $\xi_+$ is for a long localization length and $\xi_-$ is for a shorter one with the constant wave vector $q=\pi$, while for the latter case, the localization length follows Eq. (\ref{eq:2log}).
There are no edge states when $t_{1} > t_{c}$. For the negative oblique unit cell, otherwise, we have localized edge states with finite-valued localization lengths $1/\abs{\log({t_{2}})}$ corresponding to winding numbers at $t_1 = 0$, as shown in Fig. \ref{fig4} (f). This model is the same as the SSH model with intra- and inter-cell hopping amplitudes $d$ and $t_{2}$, respectively.

\section{Discussion and conclusion}

A generalized bowtie ladder in which the cross-chain hoppings are asymmetric can be decomposed into two SSH chains through direct inter-chain coupling $d$ with physical properties equivalent to the vertical bowtie ladder. Its rich phase diagram contains different phases from which a quantum phase transition through the gap closing can be distinguished. The phase boundaries are determined by the zero points of the spectrum, $(t_1+t_2)\cos{k}+1=0$ and $(t_1-t_2)\sin{k}=0$, which means that the bulk gap is closed at the transition point between different insulators possessing different winding numbers~\cite{Thon06,Li17}. It is known that the quantum phase diagram is characterized by the edge states in finite systems, as well.
For the diagonal line of the quantum phase diagram in Fig.~\ref{fig3}, there are metal-insulator transitions at $t_1=t_2=0.5$. The gap closing line $t_1=t_2>0.5$ separates two kinds of insulators possessing opposite winding numbers. The off-diagonal line $t_2=1-t_1$ provides a normal/non-trivial insulator transition.

On the other hand, we can also choose two kinds of oblique unit cells for the generalized bowtie ladder: negatively and positively rotated oblique cells, for which the next nearest couplings are $t_1$ and $t_2$, respectively. 
It is notable that the oblique bowtie ladder is smoothly transformed into an extended SSH chain with asymmetric next nearest coupling. The extended SSH chain is a simple analogue to the long-distant Haldane model that possesses a higher-order Chern number phase~\cite{Doru13, Lin15}. We treat the phase factor of the next nearest neighbor hopping to be double compared to the nearest hopping. The momentum vector doubling increases the winding number and the number of localized states due to the same topological origin of the higher-order Chern number phase. 

In conclusion, we studied the topological zero-energy edge states in bowtie ladders with different cutting edges corresponding to the choice of unit cell. The orientations of the unit cells designate different winding numbers as the topological invariant such that a bowtie ladder with a non-zero winding number contains localized edge states. We can understand the localized edge states of the general bowtie ladder models by means of the continuous transformation to coupled SSH and extended SSH. Moreover, this work reports that, compared to SSH models, the general bowtie ladder models show richer phase diagrams characterized by a bifurcation of the localization lengths of the edge states. Ultimately, this approach offers intuitive features to extract the topological invariant of a lattice model.

\section*{Acknowledgments}

This work was supported by Project Code (IBS-R024-D1), a National Research Foundation of Korea (NRF) grant (NRF-2019R1F1A1051215), and the Korea Institute for Advanced Study (KIAS) funded by the Korean government.


\begin{thebibliography}{150}

\bibitem{Hasan} Hasan, M. Z. \& Kane, C. L., {\it Colloquium}: Topological insulators, Rev. Mod. Phys. {\bf 82}, 3045 (2010).

\bibitem{Xiao} Qi, Xiao-Liang \& Zhang, Shou-Cheng, Topological insulators and superconductors, Rev. Mod. Phys. {\bf 83}, 1057 (2011).

\bibitem{Fradkin}  Fradkin, Eduardo, Field Theories of Condensed Matter Physics 2nd Ed. (Addison-Wesley Publishing, Reading, MA, 2013).

\bibitem{Del11} Delplace, Ullmo, P., D. \& Montambaux G., "Zak phase and the existence of edge states in graphene," Phys. Rev. B {\bf 84}, 195452 (2011).

\bibitem{Asb16} Asb{\'o}th, J. K., Oroszl{\'a}any, L. \& P{\'a}lyi, A., {\it A Short Course on Topological Insulators: Band Structure and Edge States in One and Two Dimensions, Lecture Notes in Physics} (Springer International Publishing, Switzerland, 2016).

\bibitem{Schnyder} Schnyder, Andreas P., Ryu, Shinsei, Furusaki, Akira \& Ludwig, Andreas W. W., Classification of topological insulators and superconductors in three spatial dimensions, Phys. Rev. B {\bf 78}, 195125 (2008).

\bibitem{Xiao08} Qi, Xiao-Liang, Hughes, Taylor L. \& Zhang, Shou-Cheng, Topological field theory of time-reversal invariant insulators, Phys. Rev. B, {\bf 78}, 195424 (2008).

\bibitem{Ryu02} Ryu, S. \& Hatsugai, Y., "Topological Origin of Zero-Energy Edge States in Particle-Hole Symmetric Systems," Phys. Rev. Lett., {\bf 89}, 077002 (2002).

\bibitem{Mer76} Mermin, N. D. \& Ho, T. L., Circulation and Angular Momentum in the A Phase of Superfluid Helium-3, Phys. Rev. Lett., {\bf 36}, 594 (1976).
\bibitem{Zak89} Zak, J., Berry’s phase for energy bands in solids, Phys. Rev. Lett., {\bf 62}, 2747, (1989).
\bibitem{King93} King-Smith, R. D. \& Vanderbilt, David, Theory of polarization of crystalline solids, Phys. Rev. B {\bf 47}, 1651(R) (1993).

\bibitem{Su79} Su, W. P., Schrieffer, J. R. \& Heeger, A. J., Solitons in Polyacetylene, Phys. Rev.Lett., {\bf 42}, 1698 (1979).

\bibitem{Stone84} Stone, Michael, Zero modes, boundary conditions and anomaliies on the lattice and in the continuum, Ann. of Phys. {\bf 155}, 56 (1984).

\bibitem{Ata13} M. Atala {\it et al.} Direct measurement of the Zak phase in topological Bloch bands, Nature Physics, {\bf 9}, 795-800 (2013).
\bibitem{Jun17} Rhim, Jun-Won, Behrends, Jan \& Jens, Bardarson, H., Bulk-boundary correspondence from the intercellular Zak phase, Phys. Rev. B {\bf 95}, 035421 (2017).

\bibitem{Jun18} Rhim, Jun-Won, Bardarson, Jens, H. \& Slager, Robert-Jan, Unified bulk-boundary correspondence for band insulators, Phys. Rev. B {\bf 97}, 115143 (2018).

\bibitem{Han19} Chen, Han-Ting, Chang, Chia-Hsun \& Kao, Hsien-chung, The Zak phase and Winding number, arXiv:1908.06700 (2019).

\bibitem{Lee18} Yea-Lee Lee {\it et al.} Topological Phases in Cove-Edged and Chevron Graphene Nanoribbons: Geometric Structures, $\mathbb{Z}_2$ Invariants, and Junction States, Nano Lett. 18, 7247 (2018);

\bibitem{Li17} Li, C., Lin, S., Zhang, G. \& Song, Z., Topological nodal points in two coupled Su-Schrieffer-Heeger chains, Phys. Rev. B {\bf 96}, 125418 (2017).

\bibitem{Car19} Velasco, C. G. \& Paredes, B., Classification of topological ladder models, arXiv:1907.11460 (2019).

\bibitem{Bin08} Zhou, Bin {\it et al.}, Finite size effects on helical edge states in a quantum spin-Hall system, Phys. Rev. Lett., {\bf 101}, 246807 (2008).

\bibitem{Xiao14} Dang, Xiaoqian{\it et al.} Complex band structure of topologically protected edge states, Phys. Rev. B, {\bf 90}, 155307 (2014).

\bibitem{Chao19} Li, Chao \& Miroshnichenko, Andrey E., Extended SSH model: Non-local couplings and non-monotonous edge states, Physics, {\bf 1}, 2 (2019).
\bibitem{Lix01} He, Lixin and Vanderbilt, David, Exponential decay properties of wannier functions and related quantities, Phys. Rev. Lett. {\bf 86}, 5341 (2001).

\bibitem{Zha15} Zhang, G. \& Song, Z., Topological Characterization of Extended Quantum Ising Models, Phys. Rev. Lett. {\bf 115}, 177204, (2015).

\bibitem{Thon06} Thonhauser, T. \& Vanderbilt, D., Insulator/Chern-insulator transition in the Haldane model, Phys. Rev. B, {\bf 74}, 235111 (2006).

\bibitem{Doru13} Sticlet, D. \& Pi{\'e}chon, F., Distant-neighbor hopping in graphene and Haldane models, Phys. Rev. B, {\bf 87}, 115402 (2013).
\bibitem{Lin15} Li, Linhu, Yang, Chao \& Chen, Shu, Winding numbers of phase transition points for one-dimensional topological systems, Euro. Phys. Lett. {\bf 112}, 10004 (2015).


\end{thebibliography}
\end{document}